\documentclass{article}

\usepackage{enumitem}
\usepackage{authblk}
\usepackage[
  backend=biber,
  style=authoryear,
  natbib=true,
  maxcitenames=2,
  mincitenames=1,
  maxbibnames=99,
  uniquelist=false,
  uniquename=false,
  sorting=nyt
]{biblatex}
\usepackage{algorithm}
\usepackage{algpseudocode}
\usepackage[english]{babel}
\usepackage[section]{placeins}

\usepackage[letterpaper,top=2cm,bottom=2cm,left=2cm,right=2cm,marginparwidth=1.75cm]{geometry}

\usepackage{amsmath}
\usepackage{graphicx}
\usepackage[colorlinks=true, allcolors=blue]{hyperref}

\DeclareFieldFormat{citehyperref}{%
  \DeclareFieldAlias{bibhyperref}{noformat}
  \bibhyperref{#1}}

\DeclareFieldFormat{textcitehyperref}{%
  \DeclareFieldAlias{bibhyperref}{noformat}
  \bibhyperref{%
    #1%
    \ifbool{cbx:parens}
      {\bibcloseparen\global\boolfalse{cbx:parens}}
      {}}}

\savebibmacro{cite}
\savebibmacro{textcite}

\renewbibmacro*{cite}{%
  \printtext[citehyperref]{%
    \restorebibmacro{cite}%
    \usebibmacro{cite}}}

\renewbibmacro*{textcite}{%
  \ifboolexpr{
    ( not test {\iffieldundef{prenote}} and
      test {\ifnumequal{\value{citecount}}{1}} )
    or
    ( not test {\iffieldundef{postnote}} and
      test {\ifnumequal{\value{citecount}}{\value{citetotal}}} )
  }
    {\DeclareFieldAlias{textcitehyperref}{noformat}}
    {}%
  \printtext[textcitehyperref]{%
    \restorebibmacro{textcite}%
    \usebibmacro{textcite}}}

\title{Augmented Reality-based Smart Structural Health Monitoring System with Accurate 3D Model Alignment}

\author[1]{Omar Awadallah}
\author[1]{Katarina Grolinger}
\author[2]{Ayan Sadhu}

\affil[1]{Department of Electrical and Computer Engineering, Western University, London, Ontario, Canada}
\affil[2]{Department of Civil and Environmental Engineering, The Western Academy for Advanced Research, Western University, London, Ontario, Canada}

\date{}

\begin{document}
\maketitle

\begin{abstract}
Structural Health Monitoring (SHM) has become increasingly critical due to the rapid deterioration of civil infrastructure. Traditional methods involving heavy equipment are costly and time-consuming. Recent SHM approaches use advanced non-contact sensors, IoT, and Augmented Reality (AR) glasses for faster inspections and immersive experiences during inspections. However, current methods lack quantitative damage data, remote collaboration support, and accurate 3D model alignment with the real structure. Recognizing these current challenges, this paper proposes an AR-based system that integrates Building Information Modelling (BIM) visualization and follows a flexible manipulation approach of 3D holograms to improve structural condition assessments. The proposed framework utilizes the Vuforia software development toolkit to enable the automatic alignment of 3D models to the real structure, ensuring successful model alignment to assist users in accurately visualizing damage locations. The framework also enables flexible manipulation of damage locations, making it easier for users to identify multiple damage points in the 3D models. The system is validated through lab-scale and full-scale bridge use cases, with data transfer performance analyzed under 4G and 5G conditions for remote collaboration. This study demonstrates that the proposed AR-based SHM framework successfully aligns 3D models with real structures, allowing users to manually adjust models and damage locations. The experimental results confirm its feasibility for remote collaborative inspections, highlighting significant improvements with 5G networks. Nevertheless, performance under 4G remains acceptable, ensuring reliability even without 5G coverage.
\end{abstract}

\noindent{\small\textbf{Keywords:} Augmented Reality; Structural Health Monitoring; Digital Twin; Building Information Modeling; 3D Model Alignment}

\section{Introduction}

Rapid deterioration poses an increasing threat to the integrity of civil infrastructure, as evident in the American Infrastructure Report Card of 2025 \citep{asce2025reportcard}, which shows that the health of a third of the infrastructure in the USA is in a poor state. This presents a significant concern as the accelerated degradation increases the likelihood of damage going undetected until it becomes severe, potentially resulting in catastrophic failures. Urbanization and material degradation are among the key factors contributing to the deterioration of civil infrastructure \citep{zhang2021tunnel,hao2023towards}. For example, rapid population growth leads to congestion and increased traffic loads on transportation structures, accelerating their wear and tear \citep{zhou2023smartcity}. Additionally, material degradation, such as cracking or spalling in concrete, compromises the reliability of infrastructure, reducing its ability to maintain structural integrity \citep{ranyal2023pothole}. This presents a need for proactive intervention to reduce the risks to the health of civil infrastructure, ultimately contributing to effective resolutions of existing structural damages. The economic implications of infrastructure deterioration are substantial, with estimates suggesting that years of underfunding in the U.S. roadway system have led to a \$786 billion backlog in road and bridge capital needs \citep{asce2025reportcard}. This financial burden emphasizes the urgency of developing more optimized, rapid, and cost-effective inspection and maintenance strategies.

\begin{figure}[!htbp]
\centering
\includegraphics[width=0.85\linewidth]{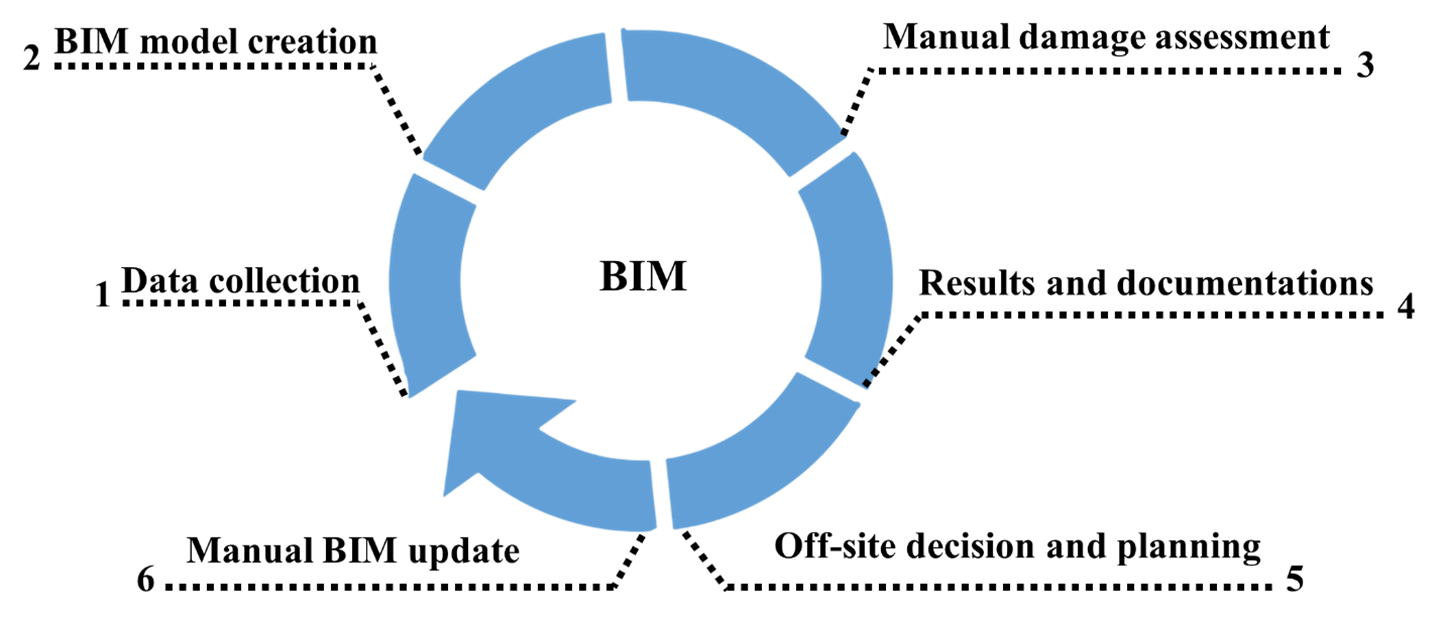}
\caption{\label{fig:fig1}Traditional flow of BIM for damage assessment.}
\end{figure}

The current conventional inspection techniques rely on periodic manual work, such as visual surveys, photographic documentation, and core sampling, with occasional non-destructive tests, which are time-consuming to set up and conduct \citep{ge2025groundrobot}. Such methods also often rely on heavy machinery to reach inaccessible areas. These can pose safety risks due to inspectors physically accessing dangerous or elevated locations \citep{mandirola2022uas,luo2023cvreview}. Such machinery also requires high operational costs, including renting, transporting, and operating it \citep{li2024digitaltwin}. In recent studies, Structural Health Monitoring (SHM) techniques have played a role in improving structural assessments, and researchers have developed Augmented Reality (AR)-based applications to modernize inspection techniques \citep{panah2021application,schiavi2022bimdataflow}. The adoption of AR technology in civil engineering has been gaining momentum, with applications extending beyond structural inspections to areas such as construction site management with AR to analyze defects \citep{awadallah2023automated}, worker safety training, and real-time visualization for structural modal identification \citep{carter2024arshm}. This broader adoption highlights the transformative potential of AR in revolutionizing various aspects of the civil engineering industry. These AR applications aid inspectors by allowing them to interact with holograms to edit or view structural data forms in real-time \citep{mascarenas2021arnextgen,morse2020interactive}. Other studies enabled inspectors to view 3D models in an AR environment beside the real structure for enhanced visualization \citep{aung2022three}. Figure \ref{fig:fig1} shows the traditional Building Information Modelling (BIM) workflow commonly adopted in structural damage assessments. Although effective, this traditional process relies heavily on manual inspections and off-site decision-making. Another framework introduced by \citet{awadallah2024remote} utilized AR technology to visualize 3D models of structures and facilitate damage assessment in augmented environments. The system utilized BIM and allowed users to load 3D models, annotate predefined damage locations using a holographic keyboard, and track data over time while supporting remote collaboration between on-site and off-site users. However, limitations exist in current studies, including the absence of quantitative data that represent damage measurements and historical records, the inability to localize damages at physical coordinates on the structure, and the lack of 3D model alignment and offset estimation with the real structure, all of which hinder accurate spatial context and comprehensive damage analysis within the framework of immersive sensing environments of AR.

This paper addresses limitations of existing literature by proposing an AR-based framework with four main contributions: (1) enabling inspectors to flexibly control and reposition 3D models within the AR environment; (2) providing correct updating and annotation of structural damage locations directly onto the 3D models; (3) automatically aligning virtual models with real-world structures using the Vuforia framework; and (4) allowing remote collaboration between on-site and off-site users for faster decision making. Building upon the previous AR system by \citet{awadallah2024remote}, this framework improves spatial context, simplifies damage assessments, and supports remote collaboration. Collectively, these contributions lead to reduced maintenance costs, prolonged structural lifespan, and enhanced public safety, aligning with sustainable urban development goals.

This study focuses on developing and validating an AR-enabled workflow that utilizes AR as the operational platform, enabling immersive, real-time monitoring directly on-site, moving beyond relying on computer systems in off-site settings. It is important to emphasize that the aim is to leverage AR’s capabilities as an enabling interface that supports SHM processes, rather than establishing AR as a superior or standalone method for SHM. Validation has been conducted on lab-scale and full-scale structures, with future work planned to extend testing across diverse real-world conditions and inspection scenarios.

The remainder of this paper is structured as follows: Section 2 discusses the related work, providing an overview of existing research and developments in the field of SHM. Section 3 delves into the proposed advanced framework, detailing its methodology, contributions, and improvements over the existing approaches. Section 4 presents the experimental results of the SHM system on two use cases: a lab-scale structure for controlled testing and a full-scale structure to assess real-world scalability. Finally, Section 5 concludes by summarizing the overall paper, the results, and potential future works.

\section{Related Work}
The recent advancements in SHM applications have integrated advanced sensor devices to improve the processing of diagnostic information for civil infrastructure \citep{kot2021recent,sakr2023iotbim}. In the majority of these studies, sensors were integrated into SHM applications to enable inspectors to process data in real-time during inspections \citep{luleci2022footbridgevr}. For example, \citet{liu2020bimmap} proposed a BIM tracker system that utilizes 3D building models to enhance the accuracy and stability of indoor localization. The system extracts information from the computer-aided design (CAD) model and categorizes it into an entity model section and a spatial element section. These sections are then processed by the proposed algorithm for map matching, aligning the 3D model with the real-world environment. The user can then view the aligned model in an AR environment using a mobile device. Despite showing promising results, the approach has multiple limitations. The system is currently limited to indoor spaces, and the alignment accuracy is limited when working in open spaces. Additionally, BIM-Tracker does not provide damage measurements, preventing the assessment of damage over time.

In another study, \citet{marino2021industry40} developed an AR-based system to assist operators in industry inspections to detect production errors by integrating BIM visualization. The system utilizes the camera of a mobile device to process live video capture of the real-world scene. The video capture is processed by the ARCore framework and scans for reference markers on the real structure. The reference markers aid the system in localizing the boundaries of the real physical structure, enabling the system to align the 3D model with the real structure. Although the system can project the 3D model onto the physical structure, its reliance on fiducial markers is impractical for real-world applications, as such markers are not typically found on public bridges and civil structures. Additionally, the system lacks numerical data for in-depth structural analysis, serving only as a visualization tool for the 3D model. Building on BIM and AR integration, \citet{garbett2021multiuser} introduced a multi-user collaborative BIM-AR system to support design and construction processes. The methodology integrates BIM with AR technologies to enhance collaboration and visualization in geographically dispersed teams. The system involved developing a mobile AR application and a large touchscreen application connected to a cloud-based database for real-time data sharing. The system allows users to view, interact with, and collaborate with 3D BIM data through the AR interface. While the study presents significant advancements in collaborative AR for construction, it acknowledges limitations, particularly the use of mobile devices for AR visualization, which limits the interactive nature of the system. This limitation could be overcome by developing systems tailored for advanced AR devices, such as the HoloLens 2 (HL2) headset.

In a BIM-centered SHM approach, \citet{fawad2023automation} automated bridge monitoring by generating FEMs and integrating sensor data within a custom IoT platform. Their research integrates BIM technology with SHM systems by developing a comprehensive methodology that includes finite element analysis of an existing bridge, generating a method for automatic finite element model generation, and linking sensors to a self-generated Internet of Things (IoT) platform programmed using the Arduino IDE. While the research demonstrates significant advancements in bridge SHM automation, the authors identify AR visualization for real-time on-site use as a future step, as the developed system is only accessible off-site. On-site monitoring is crucial for SHM as it provides in-depth data interpretation and faster decision-making on-site for engineers.

In a different study, \citet{song2023compatibility} addressed the integration of BIM models and AR to aid in semiconductor fabrication. The proposed AR-based real-time BIM data compatibility verification method aims to ensure accurate fitting between BIM models and real buildings. The study developed an AR fitting and fixing module based on Unity and the AR Foundation library, SLAM for localization, and plane detection. This system also incorporates image tracking with 2D QR markers to improve model alignment and anchoring accuracy. However, the study is tailored specifically to semiconductor fabrication plants, which possess unique characteristics such as square cube-shaped structures, potentially limiting the generalization of the findings to other types of facilities. Also, the system's reliance on 2D QR code markers placed in the environment may not be practical or feasible in all scenarios, and it does not address damage measurement and localization.

Building on the integration of advanced technologies, \citet{tan2024defectinspection} developed a comprehensive system that integrates computer vision, AR, and BIM technology for interactive inspection and defect management in construction. The proposed framework includes a computer vision model that detects damage types and quantities from real-time video capture using the YOLOv5 architecture. The system then projects the damage data in the AR environment within the inspector’s field of view, enabling on-site users to visualize damage data immediately. The last part of the system includes a BIM-based management platform for off-site users, which allows managers to visualize the BIM models alongside the damage data captured in real-time by the on-site users, enabling remote collaboration. While the system is comprehensive, the visualization of BIM models is restricted to a computer-based platform for off-site users and is not integrated into the AR scene for on-site users. Additionally, the system links defect images, classification types, and measurements without providing localization of damage locations. An improvement would involve mapping damage locations to the corresponding BIM model within the platform.

Additionally, \citet{alsabbag2024distributed} introduced the Smart Infrastructure Metaverse (SIM), a real-time distributed collaborative system for synchronous structural inspections involving on-site and remote inspectors. The methodology integrates AR and Virtual Reality (VR) technologies to enhance collaborative inspection processes. On-site inspectors use AR headsets with holographic displays to visualize digitized information overlaid on physical structures, while remote inspectors utilize VR headsets to interact with pre-built 3D point cloud maps and 360° panoramic images, generated using LiDAR sensors. The system allows both MR and VR users to perform inspection tasks such as adding annotations and taking measurements in their respective environments. While SIM demonstrates significant advancements in collaborative infrastructure inspection, it faces limitations primarily from sensor inaccuracies and drift errors in algorithm-based mapping. These issues can impact the precision of spatial alignment. However, alternative approaches such as Vuforia's Model Target feature could potentially address these concerns by eliminating the need for panoramic image stitching, avoiding drift errors, and reducing reliance on LiDAR data for spatial anchoring.

An alternative study by \citet{martins2024mrxbridge} proposed a system that integrates BIM visualization with AR through a tablet device. The system utilizes the Gamma AR framework for mobile devices to facilitate collaborative inspections. On-site users can record damage data by measuring its length and width using the AR device and storing it in an online database, enabling off-site users to access and review the information in real-time. Additionally, the system projects the 3D model onto the physical structure, enhancing visualization for on-site users. However, the researchers identified a major limitation in the accurate tracking and positioning of the model, due to the lack of depth sensors in mobile devices. This resulted in noticeable shifts from the intended location. Incorporating AR devices such as the HL2, which features depth-sensing capabilities, alongside the Vuforia software development toolkit (SDK), which utilizes depth-sensing data for 3D model alignment, could help mitigate this issue.

Focusing on more recent studies, the Vuforia SDK was integrated into the BIM-AR application to achieve more accurate 3D model alignment. \citet{fernandes2024enhancing} presented a collaborative AR platform for construction site management, integrating BIM data with real-time project information. Their system utilizes marker-based AR to align virtual models with the physical construction site, allowing multiple users to interact with and update project information simultaneously. The methodology included developing a cloud-based data management system and a mobile AR application for on-site use. However, limitations were identified in the system's need for reference markers for placement, limiting the accuracy of model alignment, as it depends on where the marker is placed with respect to the real structure. The system also relies on mobile devices, limiting the interactivity of the system.

Recently, \citet{pan2024integration} developed an AR application prototype for mobile devices using the Unity 3D engine and Vuforia SDK. This prototype overlays 3D BIM models onto real-world construction sites, facilitating improved visualization and inspection, simulating AR-BIM integration on-site with an AR app, assessing its applicability, and developing multiple-marker AR functionality. The use of multiple markers placed in various sections of the construction site aims to improve the accuracy of virtual model placement. However, the study identifies limitations, including the dependency on physical markers, which may not always be feasible in construction environments. Additionally, while the system saves images and data to an online report, this report only includes information such as the object's name, location, and description, and lacks quantitative data such as damage measurements.

In the latest study, \citet{fawad2025development} introduced the Immersive Bridge Digital Twin Platform (IBDTP), a system for automating SHM processes and enabling immersive decision-making for bridge infrastructure management. The methodology integrates Scan-to-BIM, digital twin, and AR technologies to address conventional limitations in infrastructure asset management, which has traditionally relied on static data collection. The IBDTP approach involves developing a digital twin model, which is then exported locally as an FBX format to the AR headset. While the researchers demonstrated the platform's feasibility through a case study on a single-span concrete arch bridge, the system relies on the AR headset's local storage to store 3D models, which restricts the range of models that can be utilized on-site and limits the ability to share the same model among users. Additionally, it limits the use of large models due to the headset's insufficient storage capacity for handling numerous large 3D models.

Across recent AR- and BIM-based structural inspection studies, several gaps persist: i) absence of quantitative, model-linked damage data; ii) limited real-time remote collaboration capabilities; iii) lack of automatic 3D model alignment as viewpoints change; iv) underutilization of headset interaction for field workflows; v) limited 3D model access due to local storage versus cloud-based storage; and vi) fixed model placement within AR scene. In response, this paper develops a field-ready AR inspection workflow that introduces: (i) QR-based cloud access, ensuring accurate 3D model selection and concurrent multi-user access to the same model; (ii) automatic alignment with cloud-saved annotations where new damage locations are written to the 3D model and saved for future sessions, enabling cumulative monitoring; (iii) a manual alignment mode for structures larger than the HL2 field of view or under partial occlusion; (iv) BIM-linked damage records with quantitative measurements and history, supporting progression analysis; and (v) real-time cloud synchronization, ensuring remote collaboration between on-site and off-site teams. Together, these capabilities yield inspection outputs that are repeatable across visits, comparable across users, traceable over time, and immediately actionable by distributed teams, which are critical for structural condition monitoring, asset management, and timely maintenance decisions. Table \ref{tab:related_limitations} summarizes the advantages of the proposed framework over existing studies.

\begin{table}[t]
\centering
\renewcommand{\arraystretch}{1.25}
\setlength{\tabcolsep}{4pt}
\small
\caption{Summary of advantages of the proposed framework over existing approaches.}
\label{tab:related_limitations}
\begin{tabular}{p{3.0cm} p{3.5cm} p{4.0cm} p{4.0cm}}
\hline
\textbf{Existing Related Studies} & \textbf{Key Limitations} & \textbf{Challenges} & \textbf{Advantages of the Proposed Framework} \\
\hline

Liu et al. (2020); Marino et al. (2021); Song et al. (2023); Pan and Isnaeni (2024)
&
\begin{itemize}[leftmargin=*,nosep]\setlength\itemsep{2pt}\setlength\parskip{0pt}
  \item Lack of quantitative damage measurements.
  \item No support for remote collaboration.
\end{itemize}
&
Tedious approach that involves manual data collection and delays decision-making due to the absence of real-time remote collaboration.
&
Real-time data measurements and remote collaboration enable immediate communication and decision-making.
\\
\hline

Al-Sabbag (2024); Fernandes et al. (2024); Tan et al. (2024); Fawad et al. (2025)
&
\begin{itemize}[leftmargin=*,nosep]\setlength\itemsep{2pt}\setlength\parskip{0pt}
  \item Limited integration of large-scale models within AR.
  \item Limited model alignment capability.
\end{itemize}
&
Requires heavy equipment and manual measurements for large-scale structures, increasing costs and inspection time.
&
The usage scope of the system extends to large-scale structures, overcoming cost and time challenges.
\\
\hline

Garbett et al. (2021); Fawad et al. (2023); Martins et al. (2024)
&
\begin{itemize}[leftmargin=*,nosep]\setlength\itemsep{2pt}\setlength\parskip{0pt}
  \item Underutilization of the AR headset's potential and immersive interaction capabilities.
\end{itemize}
&
High costs and safety concerns necessitate the use of heavy equipment for data collection in elevated or hard-to-access areas.
&
The use of AR headsets enables inspectors to remotely measure elevated areas from a safe and accessible location using a single equipment.
\\
\hline

Awadallah et al. (2024)
&
\begin{itemize}[leftmargin=*,nosep]\setlength\itemsep{2pt}\setlength\parskip{0pt}
  \item Fixed positioning of models.
  \item Restricted damage localization capabilities.
\end{itemize}
&
Delays in inspection due to the limited spatial context and manual localization of damages.
&
Offers improved spatial understanding with model alignment and immediate damage localization.
\\
\hline
\end{tabular}
\end{table}

To orient the reader across alignment paradigms discussed in prior work, a concise taxonomy summarized in Table \ref{tab:pose_reference_comparison} contrasts marker-based fiducials, model-target recognition, and markerless anchors based on required infrastructure and reference frames.

\begin{table}[h]
\centering
\renewcommand{\arraystretch}{1.25}
\setlength{\tabcolsep}{4pt}
\small
\caption{Summary of alignment paradigms.}
\label{tab:pose_reference_comparison}
\begin{tabular}{p{2.8cm} p{3.0cm} p{3.0cm} p{5.7cm}}
\hline
\textbf{Approach} & \textbf{Requirements} & \textbf{Pose reference} & \textbf{Advantage over Approach} \\
\hline

Marker-based
&
Printed fiducial
&
Marker-local frame
&
\begin{itemize}[leftmargin=*,nosep]\setlength\itemsep{2pt}\setlength\parskip{0pt}
  \item Avoids the need to deploy and maintain fiducial markers across large assets where placement is impractical (e.g., limited suitable surfaces on bridges).
  \item Exposure to weather or routine maintenance can damage or dislodge markers; provides an object-tied reference without on-asset alterations.
\end{itemize}
\\
\hline

Model targets
&
Prebuilt 3D model; no markers
&
Object/model frame
&
\begin{itemize}[leftmargin=*,nosep]\setlength\itemsep{2pt}\setlength\parskip{0pt}
  \item Matches the proposed system. It enables pose logging directly in the AR frame for traceable, component-specific analysis.
\end{itemize}
\\
\hline

Markerless anchors
&
No markers or CAD; device map at runtime
&
Device SLAM/world map
&
\begin{itemize}[leftmargin=*,nosep]\setlength\itemsep{0pt}\setlength\parskip{0pt}
  \item Anchors to 3D model geometry rather than a device-built map. This provides an object-fixed frame aligned with design coordinates, facilitating component-level comparisons.
\end{itemize}
\\
\hline

\end{tabular}
\end{table}

\section{Methodology}
\begin{figure}[!htbp]
\centering
\includegraphics[width=0.7\linewidth]{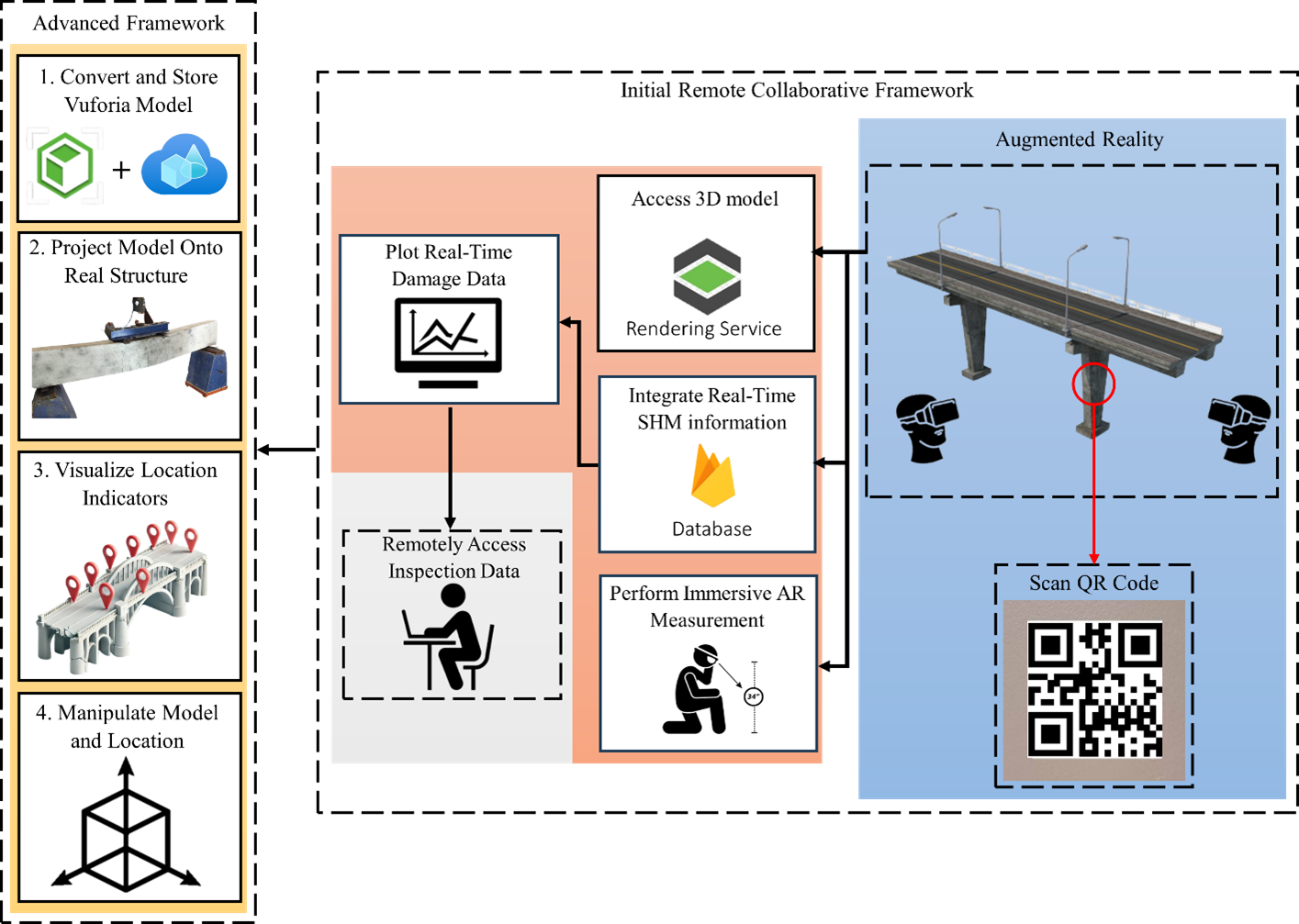}
\caption{\label{fig:fig2}Workflow of the proposed AR framework.}
\end{figure}

To overcome the limitations of previous AR-based structural inspection systems, an approach is designed to improve 3D model tracking and damage localization. The proposed approach integrates automatic model alignment with model-linked damage recording and cloud-backed collaboration. The system projects the 3D model onto the real structure, records damage at specific model elements and locations, and maintains historical records that can be revisited in later inspection sessions. Remote collaboration is supported through a shared cloud database that mirrors headset updates. Figure \ref{fig:fig2} presents the functionalities and workflows of the proposed AR framework.

\begin{figure}[!htbp]
\centering
\includegraphics[width=0.5\linewidth]{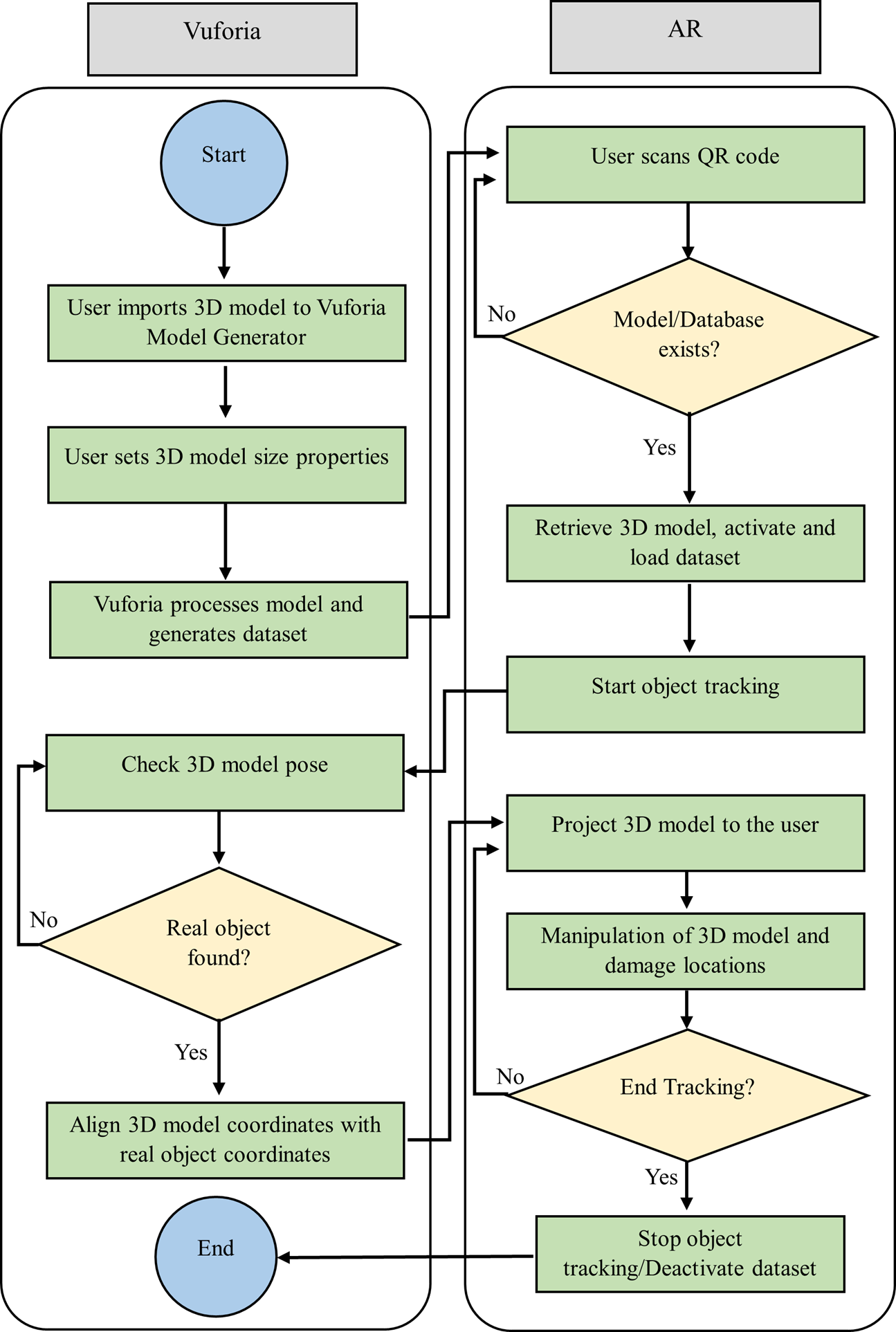}
\caption{\label{fig:fig3}Flowchart illustrating the interaction between Vuforia and AR.}
\end{figure}

In the previous study by \citet{awadallah2024remote}, the system was able to store 3D models on the cloud and render them in the AR scene. However, the system lacked alignment of 3D models onto real structures and damage localization within the 3D model itself. This prevented users from localizing the damages on the 3D model with respect to the real structure in real-time, causing off-site users to be unable to identify where the damages are localized, delaying decisions. When damages can be localized on the 3D model, and the 3D model is updated to the cloud in real-time, it can be viewed by the off-site users immediately to know where the damage is without being present on-site, enabling faster decision-making with on-site users. The Vuforia Model Target engine is introduced in this paper to aid in the automatic projection of 3D models onto real structures to visualize the correct location of damages. Vuforia’s spatial tracking accuracy on the HL2, which hosts our system, has been extensively evaluated by \citet{rieder2021hololens}, with results concluding that the HL2 delivers positioning accuracy and stability suitable for AR monitoring tasks. Unlike other tracking solutions that rely on feature points or fiducial markers, this automatic tracking approach is well-suited for infrastructure inspection, where surfaces may lack distinct textures or patterns. Figure \ref{fig:fig3} illustrates the advanced framework integrating Vuforia with the AR scene. The process begins with the import of a 3D model file into the Vuforia Model Generator, where properties such as real-world size units and spatial alignment are configured to match the actual object by specifying the scale factor and adjusting the origin point in Vuforia 3D mapping software. Vuforia then processes the features of the 3D model and automatically generates a dataset that includes the geometries of the 3D model, which is subsequently stored on the cloud. The Vuforia Model Target storage is limited to 3D models with 400,000 polygons at most (800MB), which is compatible with extensive scans of detailed civil structures (e.g., a 3D scan of a city bridge with 15M points = 430MB). Using an AR device, the user scans a QR code, which is created in advance and attached to the real structure. This QR code contains the endpoint reference to the generated dataset, enabling the AR application to connect to the database containing the 3D model and its associated dataset. The system retrieves the 3D model from the Vuforia database; these models are created in glb file format using Blender 3D graphics software. Different software, such as SolidWorks and AutoCAD, can also be used to create these 3D models. The underlying data scheme is based on polygon meshes generated from 3D point cloud scans of the structures. Each model contains vertex positions, normals, and optionally color/texture information. Once the 3D model and its geometric data are successfully loaded, the system starts the object-tracking function. The system compares the pose of the 3D model to the real environment, searching for a matching object. Upon detecting a real object, the 3D model is rendered on the HL2, and the system automatically aligns the coordinates of the 3D model with those of the real object and projects it onto the object in the AR scene in front of the user. If the object is not detected, no augmentation will be displayed, and the app will behave as if the target is absent. After the model projection, interaction with the 3D model is possible, allowing the user to manipulate the 3D model and the different damage locations in AR. This includes rotating, zooming, and repositioning the model with hand gestures if needed, as well as toggling between damage indicators displayed on a holographic panel and adding new damage indicators. When tracking ends as the user closes the application, the system stops the object tracking function and deactivates the dataset, finalizing the process.

The advanced AR framework workflow, shown in Figure~\ref{fig:fig1}, and the process between Vuforia and the AR device, shown in Figure~\ref{fig:fig3}, can be broken down into multiple detailed key steps:

\begin{enumerate}
\item \textbf{Vuforia Model Target Conversion:} To enable recognition and pose initialization in AR, the 3D model is first exported in a standard mesh format such as FBX, OBJ, or GLB and kept at true scale in meters so its size corresponds to the physical structure. The coordinate system is set with the Y-axis as the up vector and the origin positioned at the model’s center, ensuring consistent interpretation of pose during runtime. The model is then processed in the Vuforia Model Target Generator (MTG), which analyzes the mesh geometry to extract characteristic reference data. This includes descriptors of the model’s outlines, depth discontinuities, and major edge transitions that remain visible across different camera viewpoints. MTG compiles these descriptors into a Model Target dataset, which contains the geometric feature information required for reliable detection and alignment. At runtime, the proposed system utilizes this dataset to match the live camera feed with the stored reference features to initialize the model pose, after which tracking is maintained through the system. Figure~\ref{fig:mtg_conversion} shows an example of converting a sample model, for ease of illustration, using the Vuforia MTG.

\begin{figure}[!htbp]
  \centering
  \includegraphics[width=0.7\linewidth]{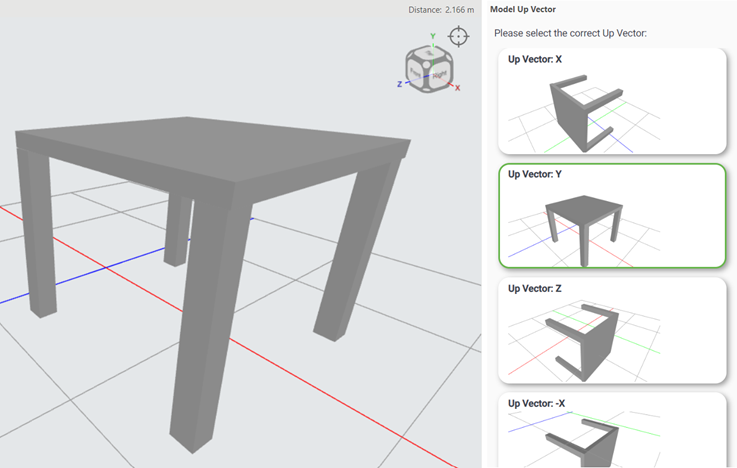}
  \caption{\label{fig:mtg_conversion}Conversion of a simple one-story 3D structural model to a compatible format using Vuforia MTG.}
\end{figure}

\item \textbf{Model Projection onto Real Structure:} In this step, the user scans a specific QR code to load the 3D model and dataset. The system loads the entire 3D model rather than only the visible parts. This design choice ensures that the full structural geometry is available for real-time condition assessment and flexible manipulation in AR, allowing users to inspect any region of the structure without restrictions. After that, the system ensures the model automatically aligns correctly with the physical structure, enabling detailed visualization of damage locations. The AR application utilizes the HL2 camera to capture the live video feed of the environment. Once the QR code is scanned, the system loads the model and the geometric features stored in the Model Target dataset. If the dataset or model corresponding to the scanned QR code does not exist, the system prompts the user to scan a different QR code. Because the user’s perspective and device position are constantly changing with head movement, the system continuously scans the environment to detect objects that match these geometric features. When a match is detected, the Vuforia Engine aligns the 3D model with the corresponding real-world object. Otherwise, no augmentation will be displayed if no matching structure is detected. This alignment process is performed globally; therefore, regions with fewer geometric features, such as the interior of box girders, will not affect the overall model alignment. Symmetries in the structure do not pose an issue because the full model geometry provides sufficient constraints for the alignment process.

\item \textbf{Location Indicator Visualization:} After projection, the next phase includes visualization of location indicators on top of the 3D model that are aligned with the location on the real structure. These indicators include a holographic panel that displays the damage information corresponding to it. When the user toggles between locations using the holographic panel shown in Figure~\ref{fig:indicators}, an animation is triggered for the specific indicator that distinguishes it from stationary indicators, highlighting for the user the exact damaged location. The holographic panel also includes other menu selections to change between measurement modes, compute measurements, and open a web application using holographic cursor clicks. This web application enables on-site and off-site users to analyze updated damage information in real-time. On-site users can view plots directly on the HL2, while off-site users can access them through a web browser on any internet-connected device. Integrated with a database and plotting engine, it allows inspectors to collaborate and track the length, area, and perimeter of damage in different locations over time. Figure~\ref{fig:indicators} shows an example of a 3D model with two damage indicators.

\begin{figure}[!htbp]
  \centering
  \includegraphics[width=0.7\linewidth]{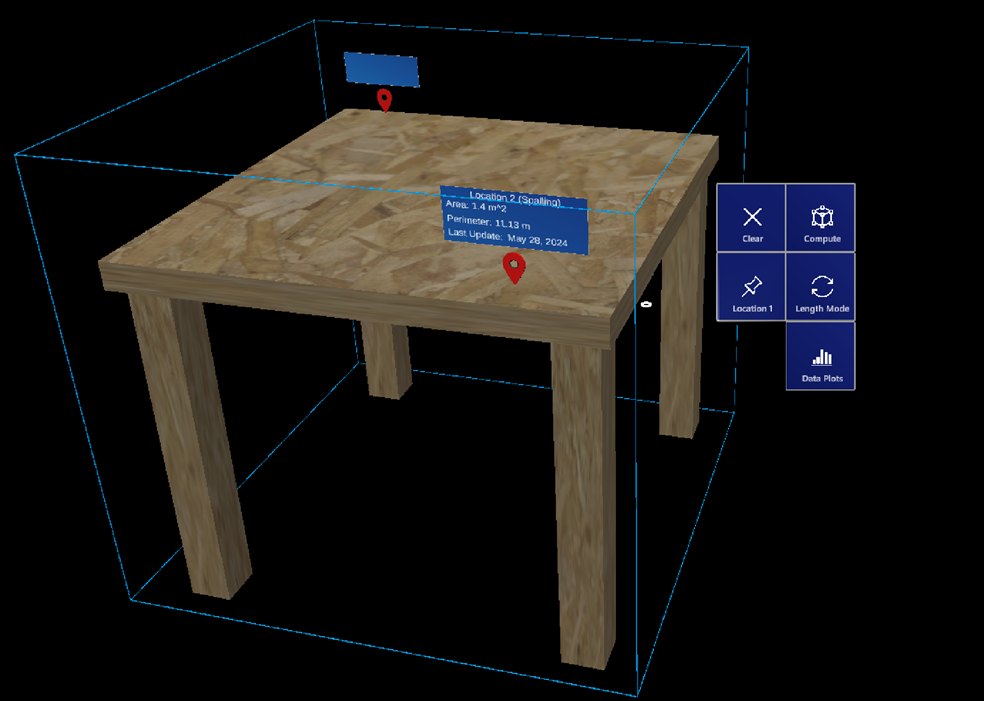}
  \caption{\label{fig:indicators}A 3D model with location indicators.}
\end{figure}

\item \textbf{Model and Location Manipulation:} This includes manual rotation, zooming, and moving the model to view it from different angles by controlling the edges of the bounding box (shown in Figure~\ref{fig:indicators}) surrounding the model using hand gestures. When a user wants to rotate the model, they can perform a twisting motion with their hands while pinching the edges of the bounding box, as seen in Figure~\ref{fig:rotate_gesture}. To zoom in or out, users can move their hands closer together or further apart while maintaining the pinch gesture. Moving the model involves grabbing the bounding box and dragging it to the desired position.

\begin{figure}[!htbp]
  \centering
  \includegraphics[width=0.7\linewidth]{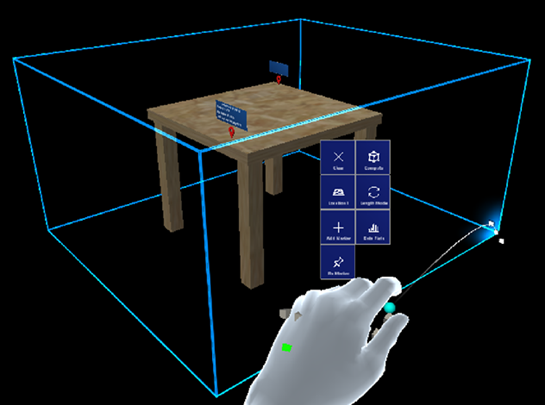}
  \caption{\label{fig:rotate_gesture}Illustration of a user rotating the 3D model.}
\end{figure}

These gestures allow for advanced manipulation of the model in the AR environment, providing a natural and immersive user experience. The user can also add new location markers as they discover new damaged locations; this procedure is explained in Algorithm~\ref{alg:damage_localization_ar}. Via the HL2, a marker is inserted and positioned on the holographic 3D model; when confirmed, its transform and metadata are saved with the model’s cloud record. In holographic applications, both the model and each location indicator are GameObjects (i.e., scene entities with a position/rotation/scale and optional components). Each new location indicator added by the user can be attached in the backend code as a child GameObject of the 3D model (parent), where they are merged into one persisted model state. On subsequent loads (any device/session), the parent model is restored with all child markers, and identified locations persist without any re-scanning. Manual manipulation offers higher flexibility to overcome automatic projection issues that may arise when the real structure exceeds the HL2 field of view (approximately $43^{\circ}$ horizontal and $29^{\circ}$ vertical), which corresponds to approximately $11.8 \times 7.8$~m at 15~m and $15.8 \times 10.3$~m at 20~m.

\begin{algorithm}[t]
\caption{Damage Localization and Annotation in AR}
\label{alg:damage_localization_ar}
\begin{algorithmic}[1]
\Require Aligned 3D Model, Existing Damage Data
\Ensure Updated Damage Annotations, Synchronized Database

\State Initialize AR environment and load aligned 3D model
\State Retrieve existing damage data from database

\While{AR session active}
  \If{user selects existing damage marker}
    \State Display damage details (type, severity, history)
    \State Allow user to edit and save updated details
  \EndIf

  \If{user initiates new damage annotation}
    \State Allow user to position damage marker using AR cursor
    \State $\mathit{selectedPosition} \gets$ AR cursor confirmed by user
    \State $\mathit{modelPosition} \gets \textsc{TransformToModelCoordinates}(\mathit{selectedPosition})$
    \State Place damage marker at $\mathit{modelPosition}$
    \State Prompt user to input damage details (type, severity, notes)
    \State Save annotation locally and upload to remote database
    \State Synchronize annotation data with remote users
  \EndIf

  \If{user adjusts 3D model position or orientation}
    \State Update model transformation matrix
    \State Recalculate positions of existing damage markers
  \EndIf

  \If{user ends AR session}
    \State Save changes and synchronize with database
    \State Exit AR environment
    \State \textbf{break}
  \EndIf
\EndWhile

\end{algorithmic}
\end{algorithm}

\end{enumerate}

Additionally, multiple headsets can join the same model session. When a user adds or edits a damage marker, the update is written to the online database with a timestamp and broadcast to all clients in real-time, including off-site users, enabling immediate decision-making. If two users edit the same item concurrently, the most recent write is kept, and the earlier version is logged for review. In summary, the automatic projection of 3D models, combined with flexible control over both the model and location markers, enables users to quickly identify damage locations during on-site inspections, ensuring productive assessment of the structure’s health.

\section{Experimental Validation}
To assess the proposed framework, the functionalities and performance of the system are tested in two practical use cases, a lab-scale experimental beam and a full-scale bridge. Beyond loading the 3D model via QR code and updating damage information, the testing focuses on various key aspects. The performance of the automatic projection of the 3D model onto the real structure using Vuforia is evaluated by visualizing how well the model aligns with the physical structure.  The system's flexibility is tested through manual manipulation of the 3D model, which involves moving, rotating, and scaling the model within the AR environment to determine how well the system handles these adjustments. The adaptability of the system is also evaluated by adding new location markers to quickly assess newly discovered damaged locations, checking how easily new markers can be integrated and visualized. Finally, the performance on network conditions is assessed by measuring the speed and reliability of data transfer and system responsiveness on 4G/5G networks by recording loading and saving times of the 3D model and damage data across 5 trials. By focusing on these specific evaluation criteria, the testing aims to examine whether the performance of the system is affected by the automatic model alignment and the damage localization features.

\subsection{Lab-scale experimental model}
\begin{figure}[!htbp]
  \centering
  \includegraphics[width=0.9\linewidth]{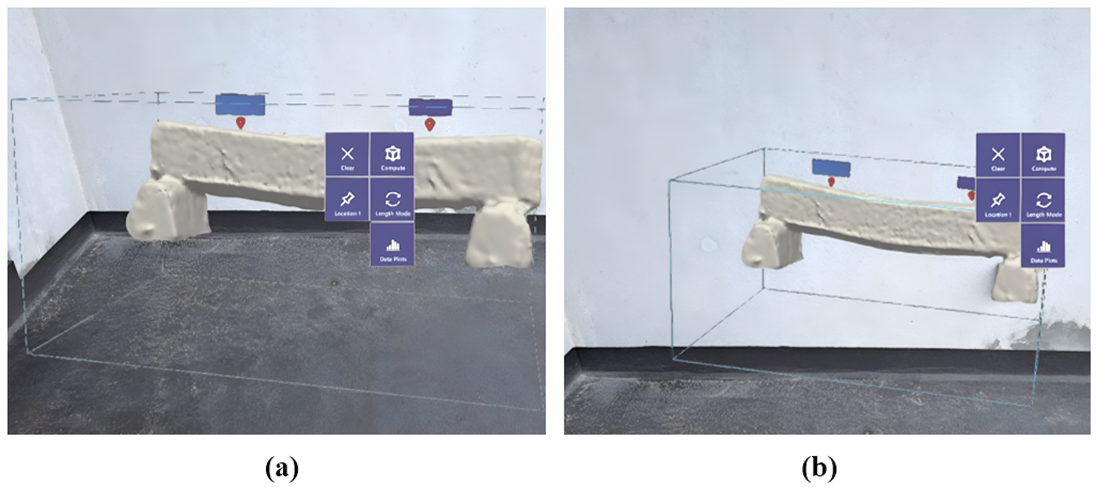}
  \caption{\label{fig:fig7}(a) Large size 3D beam model, (b) small size 3D beam model.}
\end{figure}

To evaluate the performance of the developed system, a Reinforced Concrete (RC) beam is used as a test case. A 3D model of the beam’s physical structure is created using a depth camera (ZED 2i) by capturing a live video of the entire structure and generating the model. Once generated, the model is processed using the Vuforia MTG and uploaded to the cloud database. Upon scanning the QR code, the system promptly loads the 3D model of the experimental beam. In the model manipulation step, the user is then able to manually resize the 3D model to various dimensions, demonstrating the system's responsiveness to user inputs based on the functionalities of the system. This interactive feature allows users to adjust the model size as needed by manipulating the bounding box surrounding the 3D model using hand gestures. Figure \ref{fig:fig7} illustrates this capability, showcasing the user successfully reducing the size of the model to fit within the desired parameters. This successful resizing operation showcases the flexibility of the system in accommodating user-driven modifications, enhancing its utility in practical applications.

\begin{figure}[!htbp]
  \centering
  \includegraphics[width=0.7\linewidth]{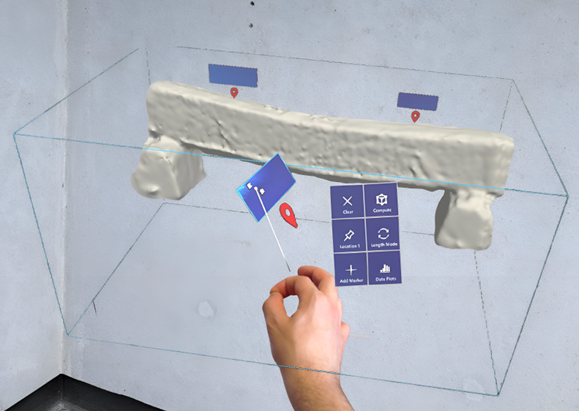}
  \caption{\label{fig:fig8}The user adding a new location marker to the 3D model.}
\end{figure}

\begin{figure}[!htbp]
  \centering
  \includegraphics[width=0.8\linewidth]{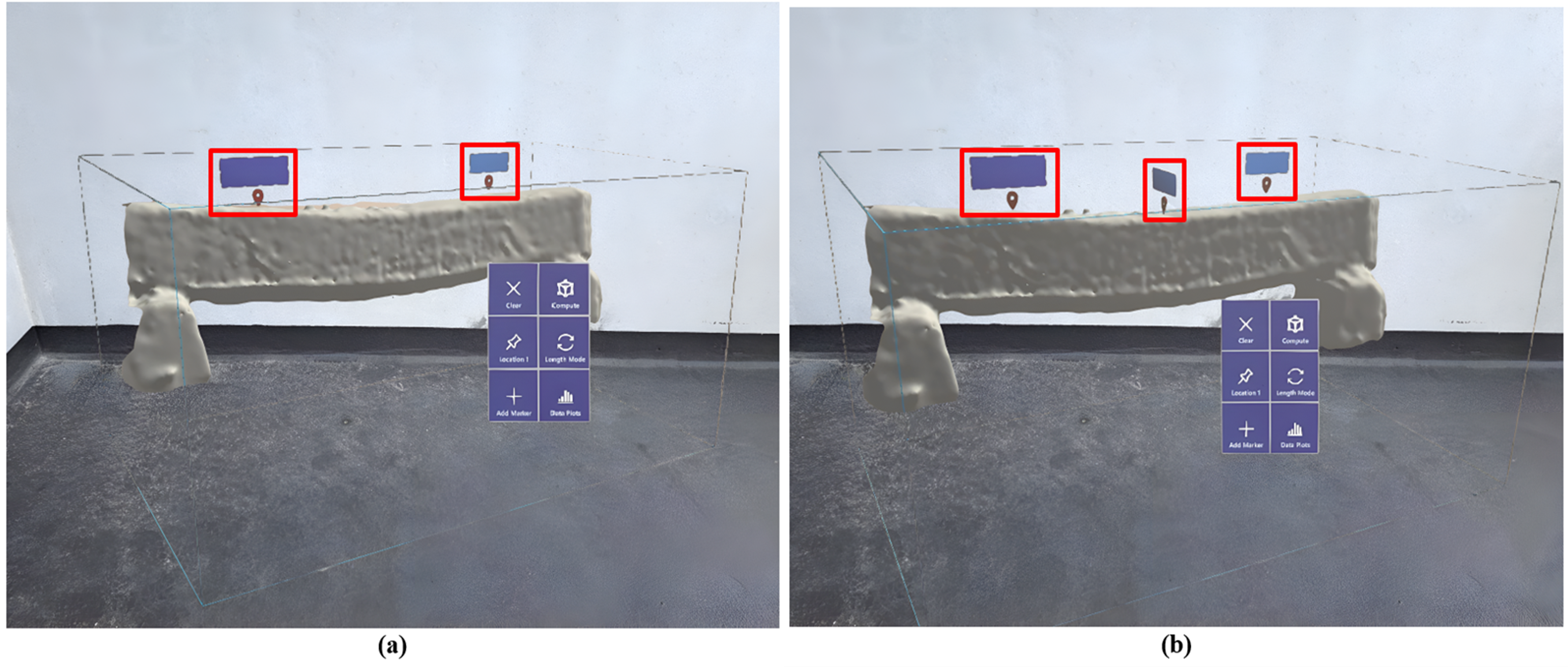}
  \caption{\label{fig:fig9}(a) Model before adding new marker, (b) model after adding new marker.}
\end{figure}

As part of the location manipulation step, the user adds a new location marker to the specific location of a newly discovered damaged location. This process involves multiple detailed steps as discussed in the third step of the framework presented in Section 3. The user accesses the holographic menu through the HL2 interface, which presents various options, including the ability to add new markers, as shown in Figure \ref{fig:fig8}. By selecting the appropriate option, the user initiates the insertion of a new marker into the AR scene. Once the marker appears in the AR environment, the user uses hand gestures to move it to the correct location of the discovered damage, on the holographic 3D model. This involves grabbing the marker with a pinch gesture, dragging it to the desired spot, and adjusting its position as needed. After positioning the marker correctly, the user performs an air-tap gesture to fix it in place. These steps facilitate the effective placement of damage markers, as demonstrated by field tests, which show that users can consistently position markers with ease in practical engineering scenarios and applications. Figure \ref{fig:fig9} shows the beam model with two location markers initially, and the beam model with three location markers after adding a new one.

\begin{figure}[!htbp]
  \centering
  \includegraphics[width=0.8\linewidth]{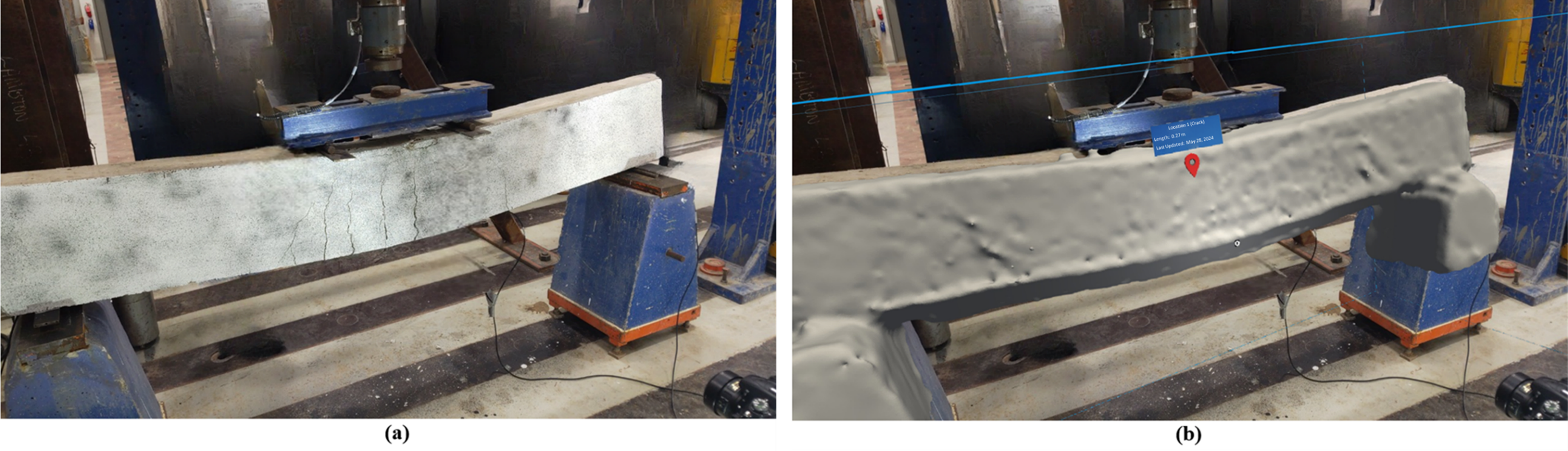}
  \caption{\label{fig:fig10}(a) Original beam structure, (b) projected 3D model with a marker indicating the location of the damage.}
\end{figure}

Finally, the 3D model is projected on top of the real beam structure, allowing the user to pinpoint the exact locations of the damage on the model and update the data. Figure \ref{fig:fig10} shows the real structure and the projected 3D model. Note that the system supports loading any 3D model, allowing models to be designed with transparency, which enables clearer visualization of damage. The system is also compatible with models of different levels of detail or texture, including highly detailed versions, depending on inspection needs.

\begin{figure}[!htbp]
  \centering
  \includegraphics[width=0.45\linewidth]{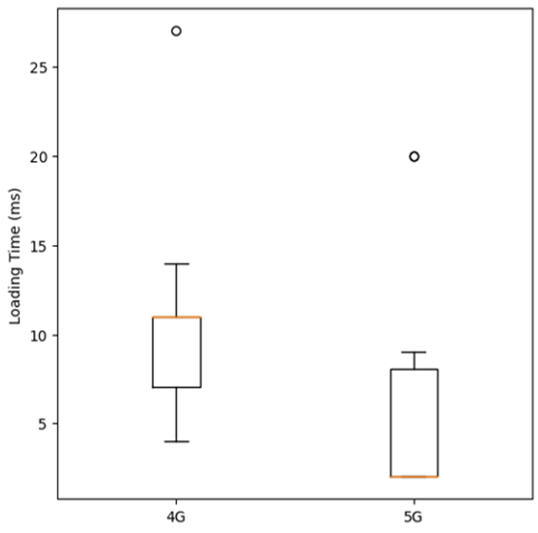}
  \caption{\label{fig:fig11}Box plot of data loading time (reinforced beam 3D model).}
\end{figure}

In the next step, the processing times of the load and save operations for the 3D model and damage data were examined to analyze whether the improved functionalities affect the overall efficiency of the system. Testing is conducted on 4G and 5G networks to evaluate their effect on remote collaboration efficiency, with processing times recorded for performance analysis. The results of loading the 3D model for the reinforced beam (~154 MB) highlight a notable speed advantage of the 5G network over 4G, with the 3D model loading in roughly half the time (~1.2s vs. ~3s). Subsequently, the loading times in milliseconds for the model's corresponding damage information are examined and depicted in the box plot in Figure \ref{fig:fig11}. The box plot reveals consistently shorter processing times for 5G compared to 4G, with the majority of 5G data showing faster load times, as indicated by the lower quartile positioning of the median line. However, the performance on the 4G network is acceptable for reliable data transmission, showing that the system is usable where the availability of the 5G network is limited.

\begin{figure}[!htbp]
  \centering
  \includegraphics[width=0.45\linewidth]{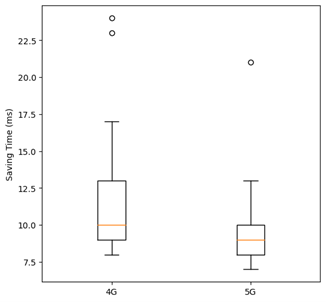}
  \caption{\label{fig:fig12}Saving times box plot of the beam damage data for the advanced system.}
\end{figure}

Finally, the box plot in Figure \ref{fig:fig12}
compares the saving times of the new damage measurements in milliseconds for 4G and 5G networks. The median saving time for 5G is slightly lower than that of 4G, indicating faster performance. Additionally, 5G shows a narrower interquartile range (IQR), suggesting more consistent saving times compared to the wider IQR and greater variability seen with 4G. The 5G network also has fewer and less pronounced outliers, further demonstrating its reliability and efficiency over 4G. On the other hand, the processing times for 4G are within 20-30 milliseconds, showing that it can be used for AR-based inspections if needed. Overall, 5G offers faster and more stable saving times than 4G, indicating that the functionalities did not affect the performance and efficiency of the system. Similar to data loading operations, the saving operations are acceptable for real-time remote collaboration between on-site and off-site users when 5G is absent, where data is saved in less than a second.

\subsection{Full-scale Bridge}
\begin{figure}[!htbp]
  \centering
  \includegraphics[width=0.7\linewidth]{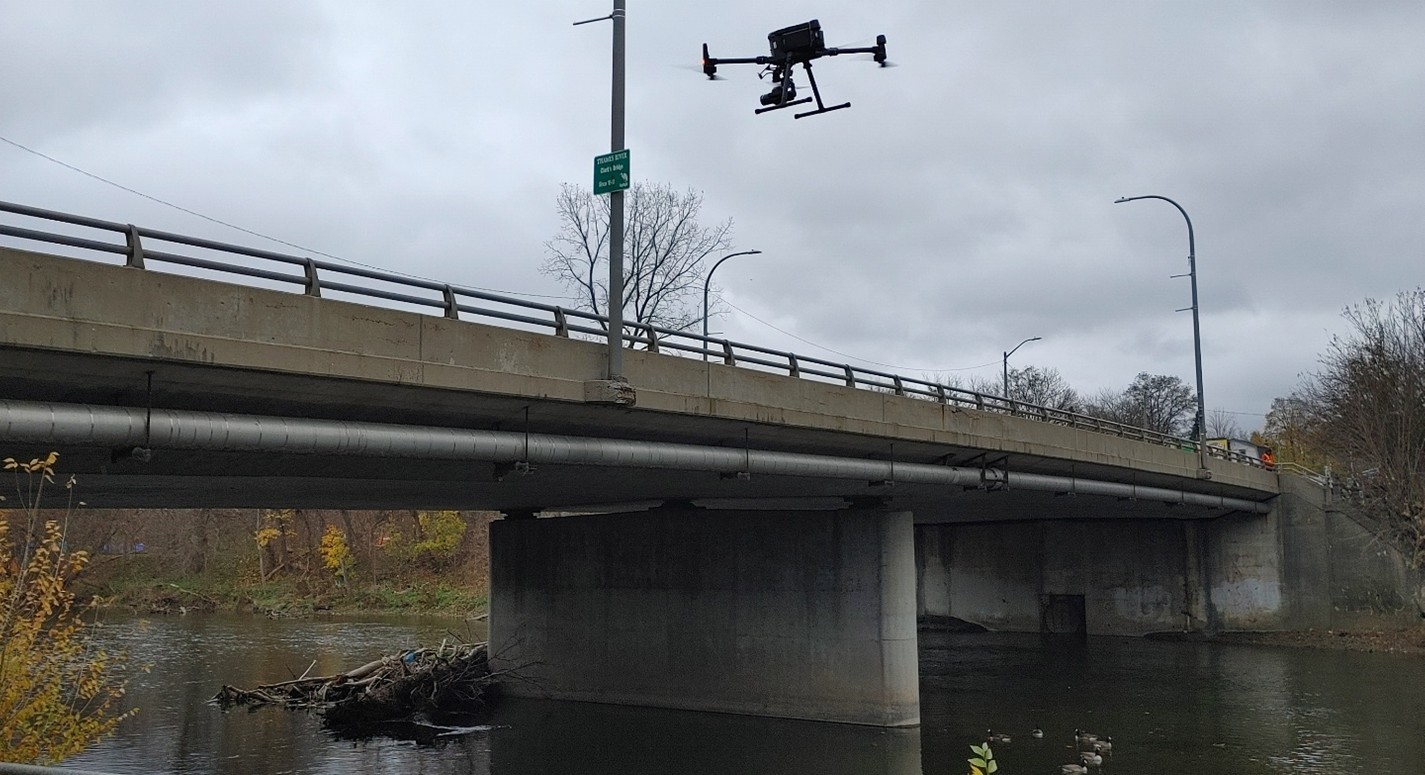}
  \caption{\label{fig:fig13}Full-scale city bridge inspected using a lidar and camera-mounted drone.}
\end{figure}

To further examine the behavior of the system in the real world, the advanced system was tested on a full-scale bridge, shown in Figure \ref{fig:fig13}, by projecting the 3D model of the bridge on the real structure. The bridge used in this study is an RC bridge located in London, Ontario, Canada.

\begin{figure}[!htbp]
  \centering
  \includegraphics[width=0.7\linewidth]{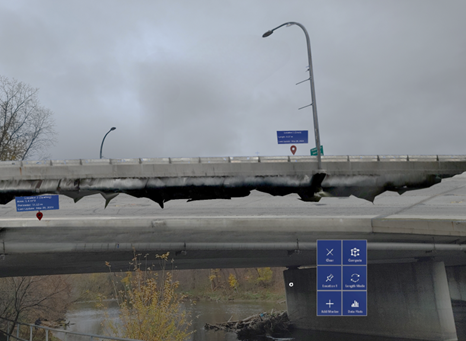}
  \caption{\label{fig:fig14}3D model projection of the bridge on the real structure.}
\end{figure}

A drone equipped with a LiDAR sensor, shown in Figure \ref{fig:fig13}, was used to scan the structure, generating four 3D mesh segments. These segments were then integrated using 3D modeling software to produce the final 3D model. Figure \ref{fig:fig14} shows the bridge after projecting the 3D model in the AR scene. In addition to 3D model projection, multiple measurements were taken for two instances of spalling and two instances of cracking. Each measurement was computed using the HL2 by placing segmentation points around the damage, mapping them to 3D real-world coordinates, and computing key metrics such as length, area, and perimeter, ensuring correct localization and data accuracy. During the inspection, each damage location was marked by the user with a location marker, ensuring that the measurements could be correctly referenced and correlated with their physical locations on the bridge, as shown in Figure \ref{fig:fig14}. The marked locations serve as reference points, allowing for easy identification and correlation with the detailed measurement data presented in the AR scene.

\begin{figure}[!htbp]
  \centering
  \includegraphics[width=0.7\linewidth]{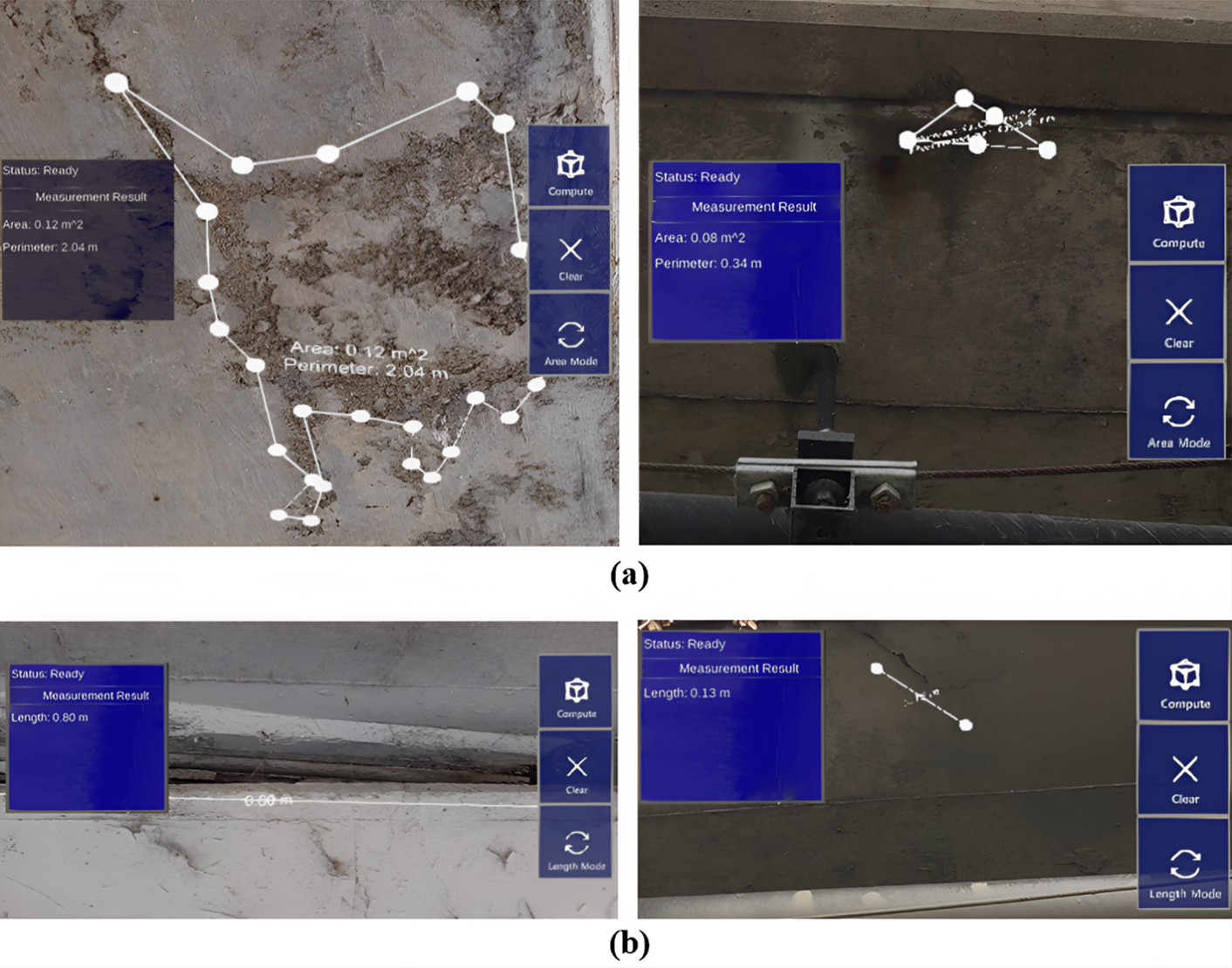}
  \caption{\label{fig:fig15}(a) HL2 spalling measurements, (b) HL2 crack measurements.}
\end{figure}

The measurement process involved various detailed steps. First, the user identified the damage locations using visual inspection and the AR overlay provided by the HL2. The user then placed location markers at each identified damage site. Once the markers were in place, the user used the measurement tool to place segmentation points around the damaged region to compute the severity of spalling and cracks. These measurements included the length of cracks and the area and perimeter of spalling as seen in Figure \ref{fig:fig15}. 

The accuracy of this measurement approach was quantitatively evaluated under realistic on-site conditions with domain experts \citep{awadallah2025arbridge}, demonstrating its reliability within practical inspection constraints. The data was then automatically saved to the remote database associated with the corresponding location markers in the 3D model. The damage data is stored in an online database in JSON file format, where each detected damage instance includes its class label, prediction score, damage measurement, and the date the analysis was performed. The damage data schema is summarized in Table \ref{tab:table3}. Damage records are maintained as time-stamped entries keyed by location ID and damage label; each new inspection appends a record (length, area, perimeter, date). A web dashboard then aligns records by location ID and timestamp to plot measurements over time, where users can visualize them. This structured format allows easy querying, visualization, and integration with structural condition assessment workflows. For example, the data can be aggregated over time or across structural components to compute condition indices such as Bridge Condition and Health indices (BCI/BHI).

\begin{table}[t]
\centering
\caption{Damage data schema.}
\label{tab:table3}
\renewcommand{\arraystretch}{1.2}
\setlength{\tabcolsep}{6pt}
\begin{tabular}{l l l}
\hline
\textbf{Field} & \textbf{Type} & \textbf{Units} \\
\hline
ID            & Integer (whole number)            & --- \\
Damage label  & String                            & --- \\
Length        & Float (continuous value)          & meters \\
Perimeter     & Float (continuous value)          & meters \\
Area          & Float (continuous value)          & m$^{2}$ \\
Date          & String                            & dd/mm/yy \\
\hline
\end{tabular}
\end{table}

The AR overlay of the measurements with the damage location markers helps users analyze the damage data at all locations. The real-time feedback from the HL2 also allows for immediate adjustments, ensuring that the data collected is accurate and reliable.

\begin{figure}[!htbp]
  \centering
  \includegraphics[width=0.5\linewidth]{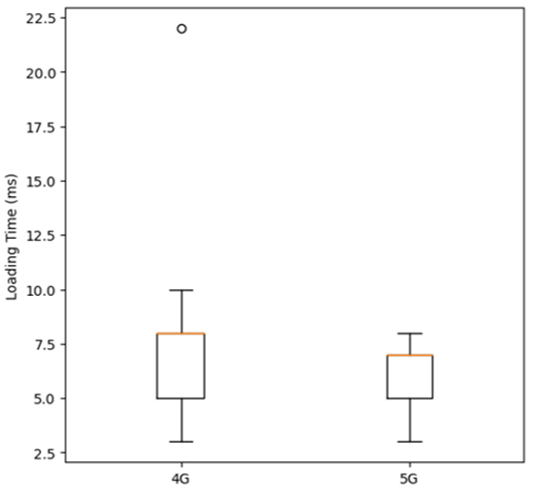}
  \caption{\label{fig:fig16}Box plot of data loading time (bridge 3D model).}
\end{figure}

Additionally, the ability to visualize and interact with the 3D model in the AR environment enhances the user's understanding of the damage and its context within the overall structure. To further assess the system's performance, the processing time for load operations of the bridge 3D model (~430 MB) was measured, and it reveals significant differences between 4G and 5G networks.  The time required to load the 3D model is drastically reduced with 5G, averaging around 1.5 seconds compared to the 4 seconds required by 4G. This decrease in loading time highlights the substantial efficiency gains provided by the 5G network, enabling faster access to critical data and reducing the waiting time for users during inspections. Similarly, the box plot for loading times of damage information in Figure \ref{fig:fig16} further supports these findings. The median loading time for 5G is slightly faster than that for 4G, and the interquartile range (IQR) for 5G is narrower. This indicates that 5G not only provides faster loading times on average but also delivers more consistent performance, with less variability in loading times. Such consistency is crucial in real-world applications where predictable system behavior can enhance user experience and reliability. On the other hand, the damage data loading times for 4G are relatively close to 5G, and when dealing with small data sizes, it shows that 4G can provide acceptable and similar performance to 5G, allowing users to work with 4G if the 5G network is limited.

\begin{figure}[!htbp]
  \centering
  \includegraphics[width=0.5\linewidth]{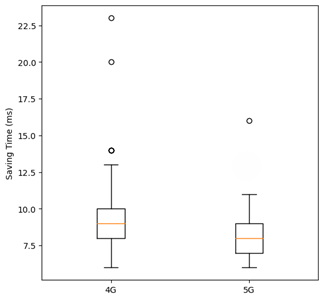}
  \caption{\label{fig:fig17}Saving times box plot of the bridge damage data for the advanced system.}
\end{figure}

The saving times, as presented in the box plot in Figure \ref{fig:fig17}, emphasize the advantage of 5G even more clearly. The median saving time for 5G is lower than that of 4G, showcasing its superior performance. Additionally, the IQR for 5G saving times is narrower, indicating less variability and more reliable performance. This means that users can expect consistently faster data-saving operations, reducing downtime and improving the overall efficiency of the inspection process. Both plots collectively demonstrate that 5G consistently delivers faster and more reliable performance across both loading and saving operations. These results confirm that the advanced functionalities of the system, such as real-time data updates and detailed damage measurements, do not negatively impact performance. Instead, the system benefits from the improved network capabilities of 5G, ensuring that it remains responsive under real-world conditions.

Finally, the implications of these findings are significant for infrastructure monitoring and maintenance. Performance testing on 4G and 5G networks revealed that 4G presented acceptable performance in a lab setting and real-world scenario, indicating that 4G can be used if access to 5G is limited. However, 5G consistently outperformed 4G, and the faster loading and saving times enabled by 5G enhance the system's responsiveness, allowing for more optimized inspections and quicker decision-making. This is particularly important in scenarios where timely updates are crucial, such as identifying and addressing structural damage to prevent further deterioration or failure. Moreover, the consistent performance provided by 5G ensures that the system can be reliably used in various locations, making it a versatile tool for civil engineers and inspectors. The integration of HL2 and 5G technologies thus represents a significant advancement in the field of SHM, offering a powerful solution for maintaining the safety and integrity of critical infrastructure.

\subsection{Registration Alignment Performance}
\begin{figure}[!htbp]
  \centering
  \includegraphics[width=0.7\linewidth]{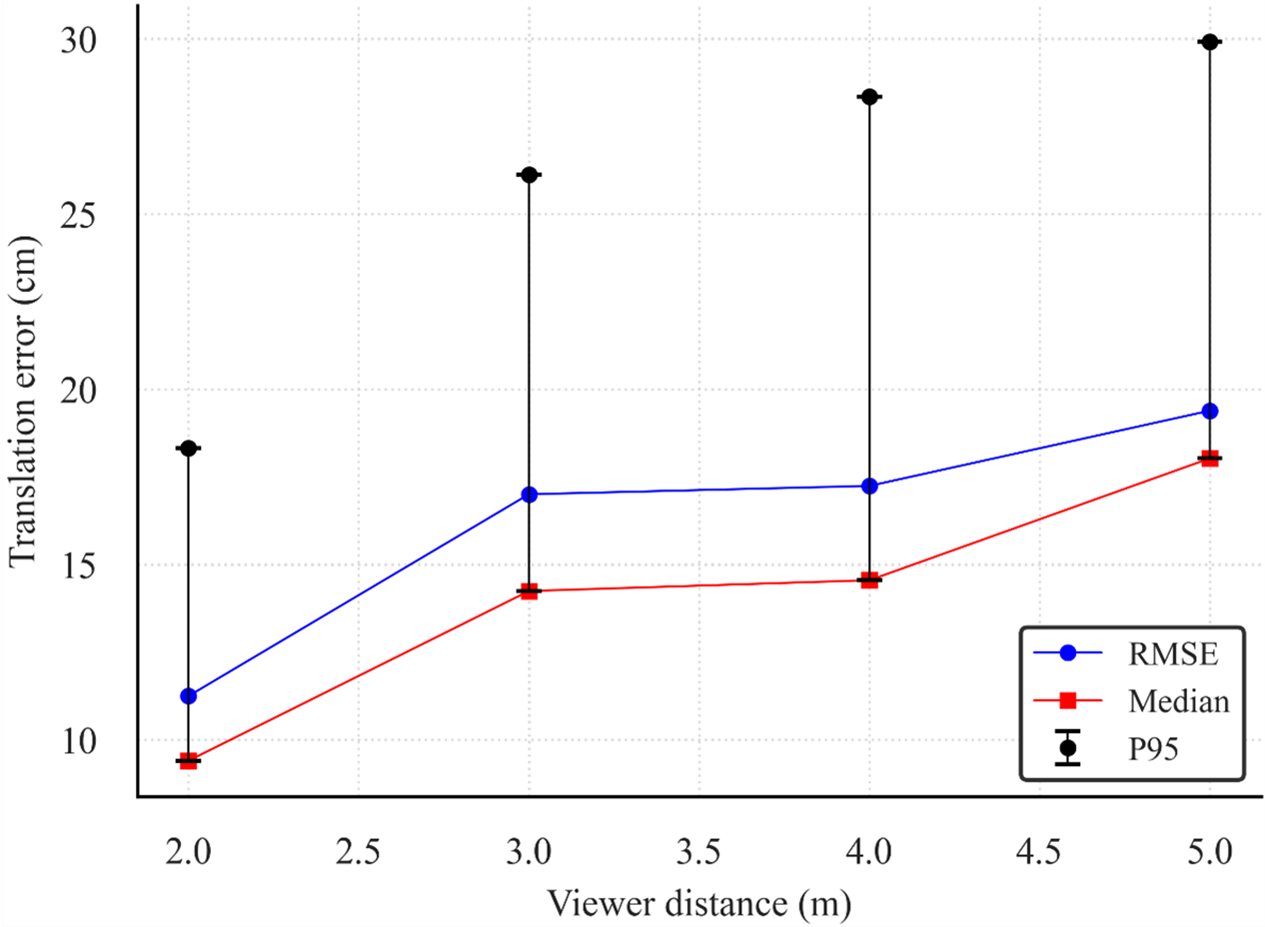}
  \caption{\label{fig:fig18}Translation RMSE versus viewing distance.}
\end{figure}

The test setup of the alignment performance was conducted on the lab-scale case study using the HL2 to overlay a 3D model onto the physical structure. The user stood in front of the structure and performed repeated placements at varying distances (2, 3, 4, and 5 meters). At each distance, the user executed 5 independent placements following the same viewing approach (initial fix, brief lateral pan, and settle), yielding a total of 20 runs. After each placement, the system recorded the 3D model and real structure poses in the same world-coordinate frame once alignment is established, providing their positions and orientations. Translation error is computed as the size of the offset between their positions, using the differences in x, y, and z. Rotation error is computed as the smallest angular turn needed to align one orientation with the other, reported in degrees. The medians (p50), 95th percentiles (p95), and root‑mean‑square error (RMSE) per distance were reported. Figure \ref{fig:fig18} summarizes how translation error scales with viewing distance. As shown, error increases as the distance between the user and the physical structure increases: the translation RMSE rises from 13.0 cm at 2 m to 19.0 cm at 5 m (intermediate values: 17.0 cm at 3 m and 4 m). This trend reflects the reduced pixel‑to‑world scale and pose uncertainty at larger distances. The median and 95th percentile shown in the figure are summarized and discussed in Table \ref{tab:tabl4}.

Table \ref{tab:tabl4} lists, for each distance, translation, and rotation statistics (median, p95, and RMSE). Translation error grows steadily with distance: the median rises from 9.40 cm (2 m) to 14.24–14.55 cm (3–4 m) and 18.03 cm (5 m), while the p95 expands from 18.33 cm to 26.13 cm, 28.35 cm, and 29.92 cm, respectively. The RMSE follows the same progression (11.25 cm → 17.01 cm → 17.24 cm → 19.39 cm), indicating that both typical and tail errors increase as the user stands farther back. Rotation shows a milder dependence on distance: median rotation error remains around 3.0–4.4°across 2–5 m, the p95 grows from 5.84° (2 m) to 8.59° (5 m), and RMSE increases from 3.62° to 5.31°. These values quantify two practical points for remote inspection workflows: (1) translation accuracy remains in the centimeter range for the bulk of trials even at 5 m, and (2) orientation is generally stable, with high-percentile rotations staying under ~9°, which is typically manageable for visual assessment and measurement overlays.

\begin{table}[t]
\centering
\renewcommand{\arraystretch}{1}
\setlength{\tabcolsep}{5pt}
\caption{Per‑distance summary.}
\label{tab:tabl4}
\begin{tabular}{c c c c c c c}
\hline
\\[-0.9em] 
\textbf{\shortstack{Distance \\ (m)}} &
\textbf{\shortstack{Trans\_RMSE \\(cm)}} &
\textbf{\shortstack{Trans\_Median \\(cm)}} &
\textbf{\shortstack{Trans\_P95 \\(cm)}} &
\textbf{\shortstack{Rot\_RMSE \\(deg)}} &
\textbf{\shortstack{Rot\_Median \\(deg)}} &
\textbf{\shortstack{Rot\_P95 \\(deg)}} \\
\hline
2 & 11.25 & 9.40  & 18.33 & 3.62 & 3.17 & 5.84 \\
3 & 17.00 & 14.24 & 26.13 & 3.43 & 3.05 & 5.76 \\
4 & 17.24 & 14.55 & 28.35 & 4.66 & 4.41 & 6.88 \\
5 & 19.39 & 18.03 & 29.91 & 5.31 & 3.95 & 8.59 \\
\hline
\end{tabular}
\end{table}

\begin{figure}[!htbp]
  \centering
  \includegraphics[width=0.7\linewidth]{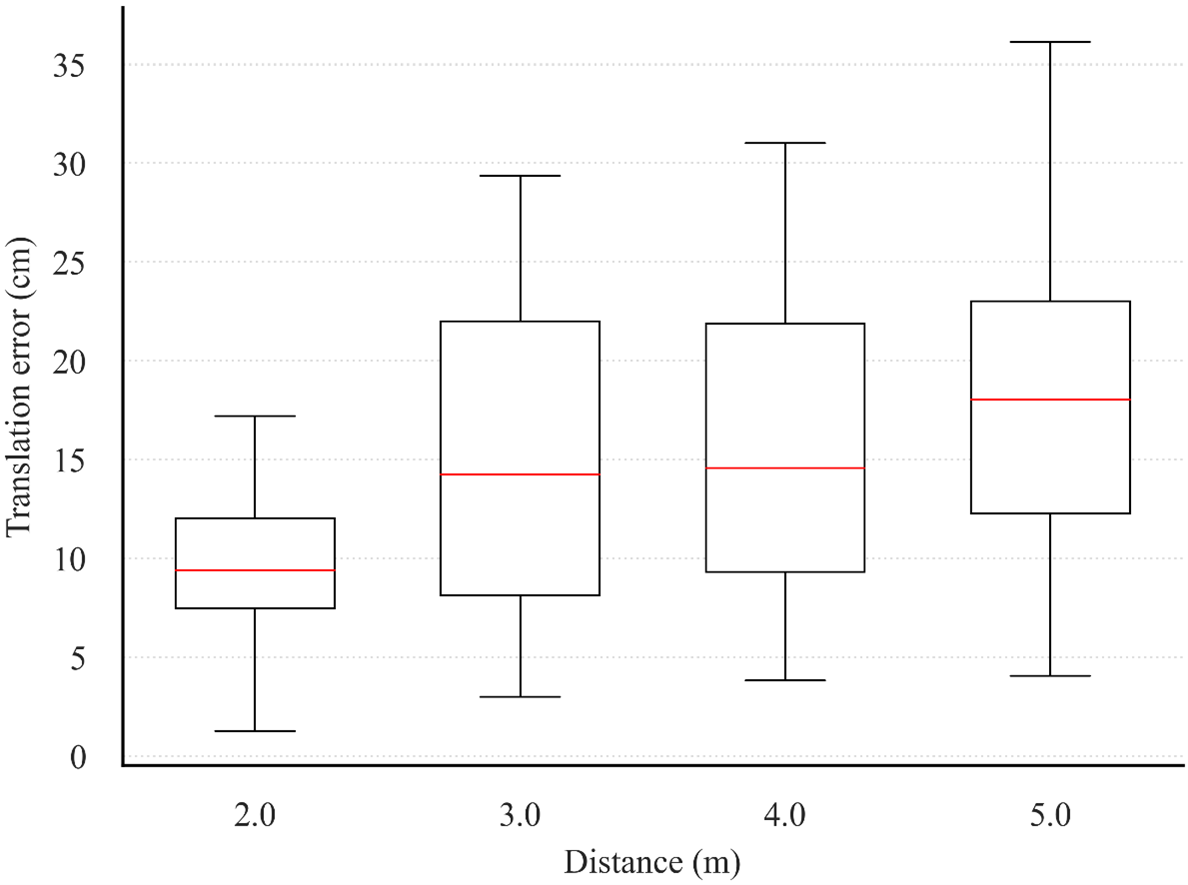}
  \caption{\label{fig:fig19}Per‑distance translation error distributions.}
\end{figure}

Figure \ref{fig:fig19} details the distribution at each distance using box plots. The central tendency increases as distance increases: the median is roughly ~10 cm at 2 m, climbs to ~14–15 cm at 3–4 m, and reaches ~18 cm at 5 m. Variability also grows with distance; the interquartile range is narrowest at 2 m and visibly widens at 3–5 m, while the upper bounds extend progressively higher, indicating a heavier high-error tail farther away. A small number of high-end outliers appear near ~19–21 cm at 2 m; however, by 5 m, the tail stretches into the mid-30 cm range. Taken together, the figure demonstrates a monotonic degradation in placement accuracy and stability as viewing distance increases. However, even with this distance dependence, the medians remain within a practical centimeter range up to 5 m, indicating the approach is suitable for remote inspection tasks where the user maintains safe stand-off while still obtaining actionable alignment accuracy.

\begin{figure}[!htbp]
  \centering
  \includegraphics[width=0.7\linewidth]{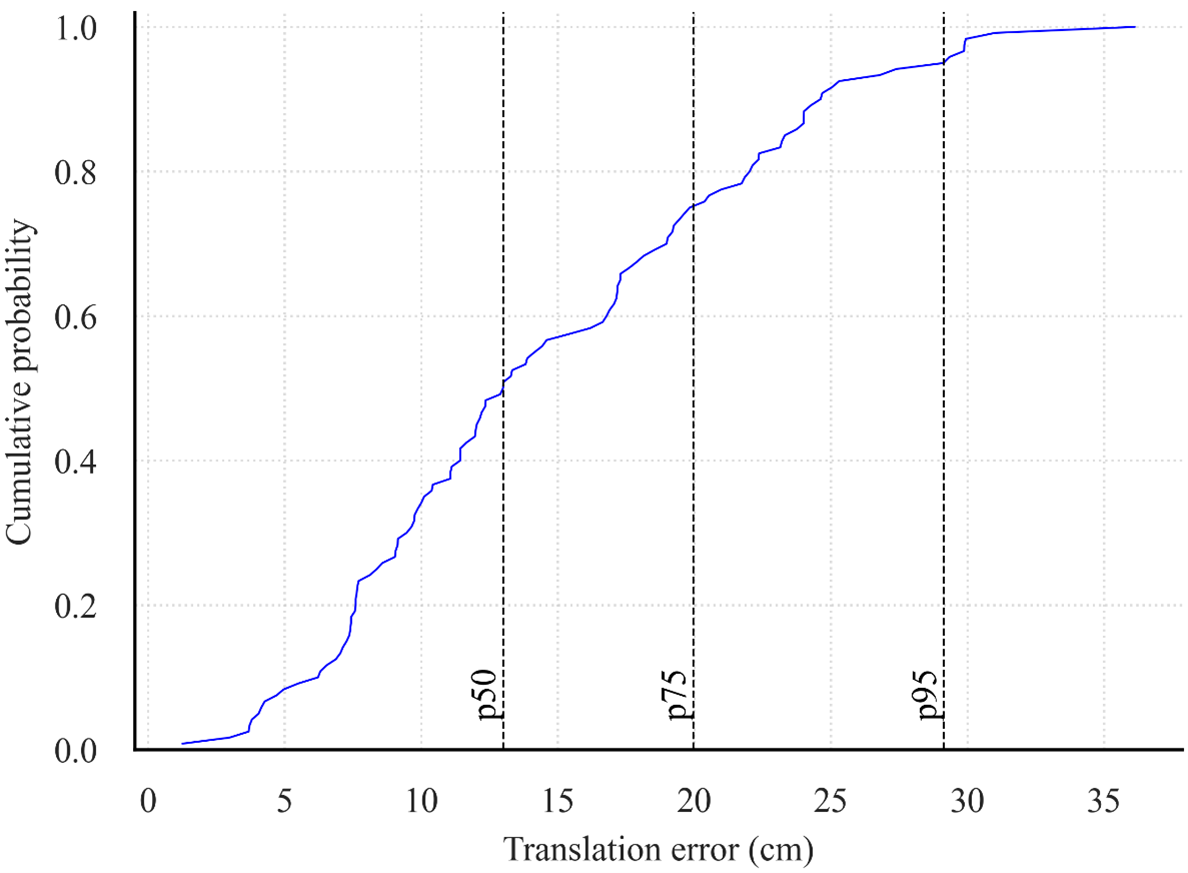}
  \caption{\label{fig:fig20}Overall CDF of translation error across all trials.}
\end{figure}

Figure \ref{fig:fig20} presents the cumulative distribution function (CDF) of translation error across all trials. The distribution indicates a median (p50) = 13 cm, p75 = 20 cm, and p95 = 28 cm, with observed errors ranging from ~3–4 cm (best cases) to ~35 cm (worst cases). Analyzing the CDF as the proportion within tolerance, a 20 cm tolerance yields an expected compliance of 75\% of trials, while 28 cm tolerance corresponds to 95\% of trials. These distributional statistics provide a clear, data-driven basis for specifying acceptance thresholds in stand-off inspection scenarios.

In summary, the presented use cases demonstrate the framework's advantages over existing methods in addressing traditional inspection challenges. The framework redefines conventional inspection practices by enabling more timely assessments through real-time remote collaboration, improved spatial alignment, and adaptability to large-scale structural applications.

\section{Conclusions}
This paper proposed a framework for real-time remote structural inspection using the HL2 AR headset and 3D models derived from BIM data. The framework provides a flexible inspection process that allows users to have control over the system, from manipulating the 3D models to controlling the damage locations. The previous studies lacked the flexibility of moving the 3D model in the AR environment by the user, could not align 3D models with the real structures, and had limited options for users to locate or add new damage locations. 
As a result, the proposed framework uses the Vuforia Model Target feature to automatically align 3D models with real structures in AR, enabling users to view accurate damage locations on the 3D model in relation to the real structure. Furthermore, the visualization phase integrates location markers with the real structure and enables users to toggle between different locations. It also allows web application access, improving usability in structural inspections. Users can add new damage locations when detected and manually adjust the 3D model (rotate, move, zoom) to adjust its placement in the AR environment, especially for large structures beyond the HL2's field of view. 
The system's functionality was tested through two use cases: a lab-scale beam and a full-scale city bridge, with damage measurements taken at different timestamps. Its performance was also evaluated on 4G and 5G networks to assess whether the system can operate in both networks. The findings conclude that 1) the system aligns 3D models with real structures in AR, enabling users to view damage locations, 2) users can manually control and adjust the 3D models and damage indicators in the AR environment, and 3) damage data transfer remains stable across 4G and 5G networks, with 5G offering faster performance, reducing 3D model loading times from 4 seconds to 1.5 seconds and data saving/loading durations by approximately 5–6 ms. In scenarios with intermittent network coverage, captured images are cached locally and automatically uploaded once connectivity is restored. Similarly, for the 3D model, the load request triggered by scanning the QR code is held until a server response is available, even under slow or unreliable network conditions. The models cannot be loaded fully offline, as they are stored in an online database, and local storage on AR headsets is not feasible due to limited storage capacity.
The study demonstrated how the advanced framework improves structural inspection using AR compared to existing work in overcoming current challenges. By projecting and aligning 3D models with real structures, users can quickly locate and assess damage. The ability to manually manipulate the 3D models and damage indicators offers greater flexibility, particularly when adjusting placement in the AR environment for large structures that extend beyond the HL2's field of view. In contrast, Model Target alignment performs best under normal lighting with an unobstructed view of distinctive geometry. Poor illumination, such as very dim light, strong backlighting, or glare, reduces edge visibility and can slow or prevent initialization. Occlusions that cover key features lower detection reliability. The current system only focuses on visible damages, which limits the assessment of any internal damages or hidden defects. Future work will: (i) explore the integration of AI for automated completion of partial or noisy scans to overcome low visibility and occlusions, (ii) investigate assessment of hidden internal damages and integration of underwater robots for below-water structures, and (iii) examine automated alignment offset correction with AI.

\section*{Acknowledgment}
The authors would like to acknowledge the City of London for its support in facilitating the field test on the full-scale bridge. The authors also thank the Ontario Graduate Scholarship for funding the first author, and the Canada Research Chairs program [grant id: CRC-2022-00261 (Ayan Sadhu) and CRC-2022-00078 (Katarina Grolinger)] for providing financial support for the proposed research, awarded to the last two authors. The authors also thank the Strategic Priority Funding of Western University for funding the Smart Cities Laboratory and the AR devices utilized in this research. Special thanks are reserved for Kyle Dunphy, Mohamed Barbosh, and Premjeet Singh (research team of the corresponding author) for their valuable assistance in the drone flight of the test bridge utilized in this paper.

\printbibliography  

\end{document}